\documentclass{article}
\usepackage{fortschritte}
\usepackage{amsmath, amssymb, amstext}

\newcommand{\cov}{\mathrm{cov}}
\newcommand{\no}{\nonumber}
\newcommand{\un}{\underline}

\begin{document}
\def\titleline{On a gauge invariant description of soliton dynamics}

\def\email_speaker{
{\tt 
bdewit,kaeppeli@phys.uu.nl
}}

\def\authors{Bernard de Wit and J\"urg K\"appeli}

\def\addresses{Institute for Theoretical Physics and Spinoza Institute,\\
  Utrecht University, The Netherlands}

\def\abstracttext{
We present important elements of a gauge and diffeomorphism invariant
formulation of the moduli space approximation to soliton dynamics. We
argue that explicit 
velocity-dependent modifications are determined entirely from gauge and
diffeomorphism invariance. We illustrate the formalism for the case of a
Yang-Mills theory on a curved spacetime background.}

\large
\makefront

\section{Introduction}

Initiated by the work of Manton \cite{Manton:1982mp}, the geodesic 
approximation for soliton dynamics and scattering has been developed and
applied in various contexts, ranging from BPS monopoles of Yang-Mills
theory, to abelian vortices, lump solutions in $CP$-models, as well as
extremal black holes. The principal idea in all these situations is to
approximate the classical dynamics of solitons by their geodesic
motion in the space of static/stationary solutions (moduli space). In
some cases, a general (albeit implicit) formula for the metric on the
moduli space was given \cite{Harvey:1993hq,Gauntlett:1994sh}. 
Such an expression is lacking for the case of gravitation, though the 
moduli space metric is known, for example, for some particular
extremal Reissner-Nordstr\"om black holes (see {\em e.g.}
\cite{Gibbons:1986cp,Ferrell:1987gf,Shiraishi:1993hz,Michelson:1999dx,Gutowski:2000ya}). 
Our interest in this question  is related to attempts to understand the
moduli space geometry for the more complicated black hole solutions
discussed in \cite{LopesCardoso:2000qm}. 

Let us first remind the reader briefly how the geodesic approximation
is derived in the simplest setting, namely for a theory without gauge
invariance.  Consider, {\em e.g.}, the Lagrangian of a non-linear
sigma model with potential,
\begin{equation}
\label{nonlinear}
  L=  \int {\rm d}^3 x \,\left({\scriptstyle{1\over 2}} g_{IJ}(\phi)
\,\partial_t\phi^I\,\partial_t\phi^J  \right) 
- V[\phi, \partial_m\phi] \,,
\end{equation}
where $m=1,2,3$ labels the spatial components (the number of spacetime
dimensions is not crucial for what follows). 
We assume that this theory has static solutions which can be
parametrized by a number of continuous integration constants, $X^a$,
which we call {\it collective coordinates}. These could, for example,
parametrize the positions of separated lumps in multi-soliton
solutions. For the purpose of this note it suffices that the solutions
can be encoded in time-independent functions $\phi^I(\vec x, X^a)$,
which characterize  completely the continuous variety of extrema of
the potential. The geodesic  approximation is effected by truncating
the fields to $\widehat\phi^I(t, \vec x)= \phi^I(\vec x, X(t))$, where
the collective coordinates can 
depend on time. The caret indicates that these are the fields that
will be reinserted into the action, obtaining an action
$S[X(t)]$ for the collective coordinates. Upon adopting
Hamilton's principle one then derives the equations of motion for the
$X^a(t)$. In the case of (\ref{nonlinear}) this yields the equations for
geodesic motion of a particle in moduli space, with corresponding metric, 
\begin{equation}
  G_{ab}(X) = \int {\rm d}^3x\; g_{IJ}(\phi(\vec x,X))\;
\partial_a\phi^I(\vec x, X)\,   \partial_b\phi^J(\vec x,X)\,.
\end{equation}
As is well-known, the symmetries of the static solutions of the
underlying field theory are reflected in 
corresponding symmetry features of the moduli space. We refrain from
elaborating on this. Because we have adopted a Lorentz frame once we
specify the static solutions, Lorentz-boosts have
no role to play in the moduli space description.

\section{Gauge theory}
In the case of a gauge theory the static solutions are in general
subject to a class of residual gauge transformations that do not
involve the time variable. This implies that these solutions are still
ambiguous and the corresponding gauge degeneracy has to be modded out when
extracting the correct moduli space description. In principle, one
could adopt a gauge condition that would result in a class of unique
solutions, depending again on collective coordinates $X^a$, which
themselves are gauge invariant. However, it is unclear 
whether this description will lead to a gauge invariant and gauge
independent moduli space metric. This question is hard to answer,
also in view of the fact that it is difficult to respect the
gauge conditions when reintroducing time. See, for instance, the
examples discussed in \cite{Harvey:1993hq,Gauntlett:1994sh}, 
where the initial gauge conditions are modified by velocity-dependent
terms. Hence we will pursue a covariant 
approach, in which none of the residual ({\em i.e.} time-independent)
gauge transformations are fixed. We will then argue that the above
mentioned velocity-dependent modifications follow from
gauge covariance and are uniquely determined within the geodesic
approximation. 

For concreteness, let us discuss a Yang-Mills theory minimally
coupled to a scalar field $\phi$ in the adjoint representation of the
gauge group, 
\begin{equation}
\label{YM-action}
  S_{\mathrm{YM}} =
 \int \mathrm{d} t\,\mathrm{d}^3 x\;
\mathrm{Tr} \Big[ -{\scriptstyle{1\over 4}} 
 F_{\mu\nu}\, F^{\mu\nu}- {\scriptstyle{1\over 2}}
 {D}_\mu\phi \, {D}^\mu\phi\Big] - V(\phi)\,.
\end{equation}
The static configurations are subject to residual gauge
transformations. Obviously, these depend on $\vec x$, but in addition
they can also depend on the collective coordinates $X^a$, so that
inequivalent solutions (characterized by different values for the
$X^a$) may be subject to different gauge transformations. This implies
that we are necessarily dealing with an extended base space
parametrized by the coordinates $(x^m,X^a)$. To define parallel
transport in this extended bundle, we need connections $(A_m,
A_a)$, where $A_a(\vec x,X)$ is a new connection field, which for the
moment is left undetermined. The Yang-Mills connections that appear in
(\ref{YM-action}) are denoted by $A_m(\vec x,X)$ and $A_t(\vec x,X)$. 
Under the residual gauge transformations with parameter
$\Lambda(\vec x,X)$, the fields transform according to,
\begin{eqnarray}
\label{eq:resgauge}
\begin{array}{rcl}
 \delta\phi &= &[ \Lambda, \phi] \,,\\
\delta A_m &=& \partial_m\Lambda -[A_m,\Lambda] \,,
\end{array}
\qquad
\begin{array}{rcl}
\delta A_t &=&  [\Lambda,A_t] \,, \\
\delta A_a &=& \partial_a\Lambda -[A_a,\Lambda] \,.
\end{array}
\end{eqnarray}
Covariant translations of the fields induced 
by shifts of the collective coordinates, take the form
\cite{Harvey:1993hq,Gauntlett:1994sh}, 
\begin{eqnarray}
\label{eq:covvargauge}
 \delta_{\cov} \phi &= &\delta X^a \,\left(\partial_a \phi -
 [A_a,\phi] \right)= \delta X^a\, D_a\phi\,,\no\\
\delta_{\cov} A_m &=& \delta X^a \,\partial_a A_m -
 D_m(\delta X^a A_a) =\delta X^a \, F_{am}\,,\no\\
\delta_{\cov} A_t &=& \delta X^a \,\left(\partial_a A_t -
 [A_a,A_t] \right)= \delta X^a \, D_a A_t \,,\no \\
\delta_{\cov} A_a &=& \delta X^b \,\partial_b A_a -
 D_a(\delta X^b A_b) =\delta X^b \, F_{ba}\,,
\end{eqnarray}
where $F_{mn}$, $F_{ma}$ and $F_{ab}$ are the nonabelian field
strengths, which are tensors in the extended space. In addition
we have $F_{mt}=-F_{tm} =D_mA_t$ and $F_{at}=-F_{ta} =D_aA_t$.

This concludes the discussion of the space of static
solutions. Subsequently we reintroduce a dependence on time through
the collective coordinates, $X^a \rightarrow X^a(t)$, which implies that
the residual gauge transformations will also become time-dependent,
$\Lambda\to \widehat\Lambda= \Lambda(\vec x,X(t))$. Note that the
transformation property for the geodesic lift of the scalar field,
$\widehat \phi = \phi(\vec x, X(t))$, remains the same, even when the
gauge parameter depends on time through $X^a(t)$. The same conclusion
holds for the other fields. However, to ensure that the (residual) gauge
invariance is maintained under these extended transformations one
needs to identify a proper connection  $\widehat{A}_t$ in order to
define a covariant time derivative. The required expression is a
modification of the original connection $A_t(\vec x,X)$ and reads 
\begin{equation}
\label{new-At}
  \widehat A_t = A_t + \dot X^a A_a\,.
\end{equation}
With this modification, $\widehat{D}_t \widehat\phi=
\partial_t\widehat\phi - [\widehat A_t, \widehat\phi]$
transforms covariantly under the gauge transformations with parameters
$\widehat\Lambda$. Note that here $\partial_t= \dot X^a
\partial_a$, as the time dependence resides in $X^a$. Hence we obtain,
\begin{equation}
\widehat{D}_t \widehat\phi= \dot X^a\, D_a \phi -[A_t,\phi]\,.
\end{equation}
The corresponding field strengths follow from  
constructing the commutators of the covariant derivatives, 
\begin{equation}
\label{f-strength-1}
\widehat{F}_{mt} = F_{mt} + F_{mb}\,\dot X^b\,,\qquad \widehat{F}_{at} =
F_{at} + F_{ab}\, \dot X^b\,.
\end{equation}
Hence the requirement of gauge invariance leads to explicit 
velocity-dependent modifications; these organize themselves into
pullback terms to the worldline in moduli space. Note that, at this
point,  there are no velocity-dependent modifications to the other
components of the gauge potentials.

Covariant translations induced by shifts $X^a(t)\to X^a(t)+\delta X^a(t)$
now involve arbitrary functions $\delta X^a(t)$ and this
aspect requires some care. Here we only note that the correct result
for the covariant translation of (\ref{new-At}),  
$\delta_{\rm cov} \widehat{A}_t= \delta X^a\widehat{F}_{at}$, does not
involve terms proportional to $\delta\dot X^a$. This result follows from
varying $X^a(t)$ in the definition of $\widehat{A}_t$ and adding a
gauge transformation with parameter $\delta X^a A_a$; the
variation is consistent with the generalized Leibniz rule,
$\delta_{\rm cov} (\widehat{D}_t\widehat\phi) = \widehat{D}_t
(\delta_{\rm cov} \widehat\phi) - (\delta_{\rm cov} \widehat{A}_t )
\widehat\phi$. This 
relation is the same as for the underlying field theory, and it is
crucial for unambiguously identifying covariant field variations
$\delta_{\rm cov}$ 
with the variations associated with the moduli
action principle. Similar results apply for the variations of the
various field  strengths.  

Replacing $F_{\mu\nu}$, $D_\mu$ and $\phi$ in (\ref{YM-action}) by
$\widehat{F}_{\mu\nu}$, $\widehat{D}_\mu$ and $\widehat\phi$,
respectively, and dropping the (constant) contribution from the
potential terms, one obtains the following moduli action, 
\begin{equation}
  S[X(t)] = \int \mathrm{d} t
  \,\left({\scriptstyle{1\over2}} G_{ab}(X)\, \dot X^a \dot X^b -
  J_a(X)\, \dot X^a \right), 
\end{equation}
where
\begin{eqnarray}\label{eq:YMeff}
G_{ab}(X) & = &   \int {\rm d}^3 x\;{\rm Tr} \Big[ F_{am}\, F_{bm} +
D_{a} \phi\, D_b \phi\Big] 
\,,\nonumber \\
J_a(X) & = &    \int {\rm d}^3 x\; {\rm Tr}\Big[ A_t\left( D_m F_{ma}
+ [\phi, D_a \phi]\right)  \Big]\,.
\end{eqnarray}
This result is invariant under residual gauge
transformations and covariant under moduli-space diffeomorphisms
(similar results were obtained in
\cite{Harvey:1993hq,Gauntlett:1994sh}). However, it still depends 
(apart from on the static solutions) on the extra connection $A_a$,
which can be eliminated in a gauge-invariant fashion by use of its
equation of motion (valid for any $\delta A_a$), 
\begin{equation}
\label{constraint}
\dot X^a \dot X^b\, \int {\rm d}^3 x\; {\rm Tr}\Big[\delta A_a\left(
D_{m} F_{mb} +  [\phi, D_b \phi] \right)\Big]  =0\,.
\end{equation}
The friction term $\int J_a \dot X^a$ did not contribute to
(\ref{constraint}), since its variation is proportional to the (static) 
$A_t$-equation of motion. Furthermore, it vanishes in the effective
action, once the constraint (\ref{constraint}) on $A_a$ is
imposed. Upon partial 
integration and comparison with (\ref{eq:covvargauge}) one observes
that (\ref{constraint}) is the well-known orthogonality condition
\cite{Manton:1982mp,Atiyah:1988jp},
\begin{equation}
\int {\rm d}^3 x\; {\rm Tr}\Big[(\delta_{\rm cov} A_m)\, (\delta_{\rm
gauge} A_m) + (\delta_{\rm cov} \phi) \, (\delta_{\rm gauge}
\phi)\Big]  =0\, , 
\end{equation}
which ensures that the geodesic motion corresponding to $\delta_{\cov}$ is
orthogonal to the gauge orbits. In this connection observe that the moduli
space metric $G_{ab}(X)$ can be written as
\begin{equation}
G_{ab}(X) =  \int {\rm d}^3 x\;{\rm Tr} \left[ \frac{\delta_{\cov} A_m}{\delta
  X^a}\,\frac{\delta_{\cov} A_m}{\delta X^b}  +\frac{\delta_{\cov} \phi}{\delta
  X^a}\,\frac{\delta_{\cov} \phi}{\delta
  X^b}\right] \,.
\end{equation}
Since the constraint (\ref{constraint}) is a covariant equation for $A_a$, we
may solve for $A_a$ and reinsert it into the expression for the metric
$G_{ab}$ without affecting gauge invariance. In principle, the above framework
can be used for more general Lagrangians, including Lagrangians that contain
terms of higher order in the field strengths.

\section{Gravitational background}
For theories invariant under spacetime diffeomorphisms no analogous
approach has been worked out so far. In this section we present the
case of a gauge theory coupled to a stationary gravitational
background, taking the residual spacetime diffeomorphisms into
account. This is a modest step towards a more complete 
treatment of theories with gravity and it will reveal the presence of
additional velocity-dependent corrections. We start from a stationary
metric in adapted coordinates, such that its components are
time independent. Together with the gauge 
fields, the metric is determined as a stationary solution of some
underlying field theory which is assumed to depend on a number of
collective coordinates $X^a$. The residual gauge transformations are
now extended to include the following residual diffeomorphisms, 
\begin{equation}
\label{res-diffs}
  t \rightarrow t + \xi^t(x, X^a)\,,\quad x^m  \rightarrow x^m +
  \xi^m(x,X),\quad X^a \rightarrow X^a + \xi^a(X)\,,
\end{equation}
where, for completeness, we also included arbitrary moduli space
diffeomorphisms. Obviously, the latter are independent of the
spacetime coordinates, so that $\partial_m\xi^a=0$. Under residual 
diffeomorphisms a scalar and the time component $A_t$ of a gauge field
transform according to\footnote{
   As before $m, n,\ldots$  denote spatial indices and $a, b,\ldots$
   label the moduli space coordinates. The spacetime indices, denoted
   by $\mu,\nu, \ldots$ comprise the spatial indices $m,n,\ldots$ and
   the time index $t$, whereas the indices $M,N,\ldots$ comprise the
   indices of the base manifold of the stationary configurations and
   thus cover both $m,n,\ldots$ and $a,b,\ldots$.}, 
\begin{equation}
  \delta_\xi \phi = - \xi^M \partial_M \phi\,, \quad  \delta_\xi A_t = -
  \xi^M \partial_M A_t\,,  
\end{equation}
whereas $A_M= (A_m, A_a)$ transforms according to 
\begin{equation}
    \delta_\xi A_M = - \xi^N \,\partial_N A_M - \partial_M \xi^N \,A_N
    -\partial_M \xi^t \,A_t \,. 
\end{equation}
Observe that the above result implies that the $A_m$, $A_a$ and $A_t$
mix in a nontrivial way. It can
be shown that this is required by the closure of the combined
algebra of residual spacetime diffeomorphisms, moduli space
diffeomorphisms and gauge transformations (always subject to the
condition $\partial_m\xi^a=0$). 

In the adapted coordinates that we use, the line element reads, 
\begin{equation}
\label{line-element}
{\rm d}s^2 = g_{tt} ({\rm d}t + \sigma_m {\rm d}x^m)^2 + g_{mn}\,{\rm
d}x^m {\rm d}x^n \,,
\end{equation}
where $g_{tt}$, $\sigma_m$ and $g_{mn}$ depend on $x^m$ and $X^a$ and
transform under (\ref{res-diffs}). In particular we note the behaviour
of $\sigma_m$ under the transformations (\ref{res-diffs}), 
\begin{equation}
\delta_\xi\sigma_m =- \partial_m\xi^t - \xi^M\,\partial_M \sigma_m
-\partial_m\xi^n\,\sigma_n\,,
\end{equation}
so that $\sigma_m$ transforms as a gauge field with respect to
$\xi^t$-transformations. Just as in the gauge theory case, we must
introduce extra connection components in order to define parallel
transport in the bundle over the extended base space parametrized  by
$(x^m,X^a)$. These extra 
fields are denoted by $\sigma_a$ and $V_a{}^m$ and are associated with
the transformation parameters $\xi^t$ and $\xi^m$, respectively. Under
(\ref{res-diffs}) they transform according to  
\begin{eqnarray}
  \delta_\xi \sigma_a &=& -\partial_a \xi^t - \xi^M\, \partial_M
  \sigma_a -\partial_a \xi^M \,\sigma_M 
  \,,\no \\
\delta_\xi V_a{}^m &=& - \partial_a \xi^m - \xi^M\,\partial_M
  V_a{}^m - \partial_a \xi^b \, V_b{}^m + V_a{}^n \,\partial_n \xi^m\,.
\end{eqnarray}
These new fields and their transformation rules have an elegant
geometrical interpretation in terms of an extended block-triangular
vielbein field,  
\begin{equation}
\label{vielbein}
  E_{\un \Omega}{}^{\Xi} = \begin{pmatrix} e_{\un \mu}{}^\nu 
  & \varnothing \cr \noalign{\vskip3mm}
   e_{\un a}{}^\nu & e_{\un a}{}^b \end{pmatrix}\,,
\end{equation}
where $\Xi= \nu,b$,  and the underlined indices refer to the
corresponding tangent space. Here $e_{\un \mu}{}^\nu$ is the spacetime
vielbein, such that  $e_{\un \mu}{}^\mu \,e^{\un \mu\,\nu}$ equals 
the inverse spacetime metric corresponding to the stationary line
element (\ref{line-element}), and
$e_{\un a}{}^b(X)$ is some reference vielbein in moduli space; the
off-diagonal block contains the new fields $\sigma_a$ and  
$V_a{}^m$, 
\begin{equation}
e_{\un a}{}^t = - e_{\un a}{}^b(\sigma_b - V_b{}^n\, \sigma_n)\,,
\qquad 
e_{\un a}{}^m = - e_{\un a}{}^b \,V_b{}^m\,,
\end{equation}
With the exception of  $e_{\un a}{}^b(X)$ all components of the
vielbein depend on both $x^m$ and $X^a$. This functional dependence
is preserved by the residual coordinate transformations
(\ref{res-diffs}) owing to the block-triangular form of the
vielbein. The tangent space rotations acting on the vielbein decompose
into local Lorentz transformations, which may depend on both
$x^m$ and $X^a$, and $X^a$-dependent orthogonal transformations of the
moduli tangent space. It is often 
convenient to impose a gauge choice on the spacetime vielbein
$e_{\un \mu}{}^\nu$, but this is not needed below. 

The covariant translations induced by shifts of the moduli now include 
the residual gauge transformations and diffeomorphisms with
field-dependent parameters, analogous to (\ref{eq:covvargauge}). For a
scalar field we thus obtain
\begin{equation}
\label{d-cov-phi}
  \delta_{\cov} \phi = \delta X^a
  \Big[D_a - \sigma_a D_t - V_{a}{}^n (D_n- \sigma_n D_t)\Big] 
   \phi\,,
\end{equation}
where the covariant derivatives contain the gauge connections. The
covariant time derivative is just given by $D_t\phi= -[A_t,\phi]$,
because there is no dependence on time. This result takes the form of
a linear combination of field-dependent diffeomorphisms and gauge
transformations. The correctness of this
formula can be verified by requiring that $\delta_{\rm cov}\phi$
transforms precisely as $\phi$ itself, using the various
transformation rules given above. Making use of the extended vielbein,
the result (\ref{d-cov-phi}) can be written as follows, 
\begin{equation}
  \delta_{\cov} \phi = \delta X^{\un a}\, D_{\un a} \phi \,,
\end{equation}
which makes it obvious that (\ref{d-cov-phi}) has the required
properties as we have expressed the result in terms of a tangent-space
derivative. 

Naturally, this result can be extended to other fields, but in those
cases one may need (dependent) spin and affine
connections. There is no obstacle for doing this, but we prefer not to
enter into the details of their construction here. Apart from this
extension we have dealt with the stationary solutions and the
structure of the corresponding moduli space. 

Reintroducing time by letting the collective coordinates become time
dependent, proceeds in the same way as for the gauge theory, except
that matters are rather more subtle. Knowing that the time component
$A_t$ of the gauge field does acquire a velocity-dependent term, one must
introduce the following velocity-dependent modifications for all the
gauge field components in order to uniformly preserve
the transformation rules, 
\begin{equation}
\label{hat-connection}
  \widehat A_{t} = A_t + \dot X^{\un a} A_{\un a} \,,\quad 
  \widehat A_{m}  = A_m   +\sigma_m\,\dot X^{\un a} A_{\un a}\,,\quad 
  \widehat A_{a} = A_a   +\sigma_a\,\dot X^{\un b} A_{\un b}\,.  
\end{equation}
We emphasize that $\dot X^{\un a} A_{\un a}$ takes a complicated form,
\begin{equation}
\dot X^{\un a} A_{\un a}= \dot X^{a} \Big[A_a - \sigma_a A_t -
V_{a}{}^n (A_n- \sigma_n A_t)\Big] \,.
\end{equation}
The derivatives of $\phi$ thus have the form, 
\begin{eqnarray}
\widehat{D}_t\widehat\phi &=& \dot X^{\un a}\,D_{\un a}\phi -
[A_t,\phi] \,,\nonumber \\  
\widehat{D}_M\widehat\phi &=& D_m\phi + \sigma_M\,\dot X^{\un a} \,
D_{\un a} \phi \,.
\end{eqnarray}
The new connections (\ref{hat-connection}) transform in an unusual
fashion under gauge transformations. Indeed, the resulting geometry with the
velocity-dependent terms  is rather complicated  and involves nontrivial
torsion. There exists a well-defined tensor calculus that allows a
systematic construction of covariant quantities such as the ones
listed above. The complications are also reflected in the
field strengths. As a first step we have constructed the following expressions
for the field strengths, which transform covariantly with respect to both gauge
transformations and diffeomorphisms, and contain velocity-dependent terms, 
\begin{eqnarray}
\label{f-strength-2}
  \widehat{F}_{Mt} &=& F_{Mt} + (  F_{M\un b} -
  \sigma_M\, F_{t \un b })\dot X^{\un b}\,,\no\\
   \widehat{F}_{MN} &=& F_{MN} - 2 \,\sigma_{[M} \,
    F_{N]\un b}\,\dot X^{\un b}\,. 
\end{eqnarray}
Obviously, these results are an extension of (\ref{f-strength-1}) and include
nontrivial corrections due to the gravitational background. They will receive
appropriate additive modifications by terms which are separately consistent
with the symmetries. Their form is fixed by other requirements, to which we
have already been alluding in the text ({\em c.f.} the paragraph fol\-lowing
equation~(\ref{f-strength-1})). One is that they have a role to play in the
covariant translations $\delta_{\rm cov}$ that we have discussed before, and
another one concerns the validity of a generalized Leibniz rule. For the
purpose of this exposition we will neglect these modifications and we will
assume that (\ref{f-strength-2}) is complete; a discussion of these subtle
issues is relegated to a separate publication.

What remains is to substitute the new field strengths and the
covariant derivatives into the action (\ref{YM-action}), which is now
covariantized with respect to diffeomorphisms by including the
spacetime metric corresponding to (\ref{line-element}). In this way,
the moduli action takes the form 
\begin{equation}
  {S}[X(t)]= \int \mathrm{d} t\, \,\left({\scriptstyle{1\over 2}}
  G_{\un a\,\un b}(X) \, \dot
  X^{\un a} \dot X^{\un b} - J_{\un a}(X) \,\dot X^{\un a} \right)\,,
\end{equation}
where
\begin{equation}
  G_{\un a\, \un b} =  \int {\mathrm d}^3 x\; \sqrt{
  g /\vert g_{tt}\vert}\; \mathrm{Tr} \Big[ g^{mn}
  \left(F_{\un a m} - \sigma_m F_{\un a t} \right)   \left(F_{\un b
  n} - \sigma_n F_{\un b t} \right) + D_{\un a} \phi\, D_{\un b}\phi
  \Big]\,, 
\end{equation}
where $g_{mn}$ and $g_{tt}$ have been defined in (\ref{line-element}), $g=\det
(g_{mn})$ and $g^{mn}$ is the inverse of $g_{mn}$. Note that the
$\sigma_m$-terms appear only to build up $\xi^t$-invariant combinations and
the integral is in fact fully invariant under gauge transformations and
diffeomorphisms.  The linear term in the moduli action is analogous to the one
presented in the gauge theory case and we refrain from further comment. The
discussion of the constraints is premature in view of the fact that one should
also include the Einstein-Hilbert term.  It will be interesting to analyze the
final result of this approach in the context of the results of
\cite{Ferrell:1987gf}.
\\[2mm]
{\bf Acknowledgement} We thank Luis Alvarez-Gaum\'e, Gabriel Lopes Cardoso,
Jerome Gauntlett, Thomas Mohaupt, George Papadopoulos, and Slava Zhukov for
stimulating discussions.



\providecommand{\href}[2]{#2}\begingroup\raggedright\endgroup

\end{document}